\documentclass[prl,showpacs,twocolumn,a4paper,floatfix]{revtex4-1}

\usepackage{amsmath,amssymb}
\usepackage{graphicx}
\usepackage{dcolumn}

\begin{document}

\newcommand{\Z}{Z_{\rm eff}}
\newcommand{\Zr}{\Z^{(\rm res)}}
\newcommand{\eps}{\varepsilon}
\newcommand{\eb}{\varepsilon _b}

\title{Detection of positron-atom bound states through resonant annihilation}

\author{V. A. Dzuba}
\email[E-mail address: ]{v.dzuba@unsw.edu.au}
\author{V. V. Flambaum}
\email[E-mail address: ]{v.flambaum@unsw.edu.au}
\affiliation{School of Physics, University of New South Wales, Sydney 2052,
Australia}
\author{G. F. Gribakin}
\email[E-mail address: ]{g.gribakin@qub.ac.uk}
\affiliation{Department of Applied Mathematics and Theoretical Physics,
Queen's University, Belfast BT7 1NN, Northern Ireland, UK}

\begin{abstract}
A method is proposed for detecting positron-atom bound states by observing
enhanced positron annihilation due to electronic Feshbach resonances at
electron-volt energies. The method is applicable to a range of open-shell
transition metal atoms which are likely to bind the positron: Fe, Co, Ni, Tc,
Ru, Rh, Sn, Sb, Ta, W, Os, Ir, and Pt. Estimates of their binding energies are
provided.
\end{abstract}


\maketitle


Our analysis has identified about 25 open-shell atoms that are likely to form
bound states with the positron. We show that for many of them binding can be
detected through resonantly enhanced positron annihilation.

The existence of positron-atom bound states was predicted by many-body theory
calculations \cite{DFG95} and proved variationally \cite{RM97,SC98} more than a
decade ago. Since then positron binding energies have been calculated for
many ground-state and excited atoms: He~$2^3$S, Li, Be, Be~$2^3$P, Na, Mg, Ca,
Cu, Zn, Sr, Ag, and Cd \cite{DFG99,DFH00,MBR02,BM06,BM07,MZB08}. They range
from $\sim \!10$~meV to $\sim \!0.5$~eV.

In spite of this wealth of predictions, experimental verification of positron
binding to neutral atoms is still lacking. To observe positron-atom bound
states and measure their energies, one needs to produce sufficient numbers
of atoms in the gas phase or in a beam, but more critically, to find an
efficient way of populating these bound states. Thus, radiative
recombination, $A+e^+\rightarrow e^+A+\hbar \omega $, is inefficient because
of the small cross section, $\sigma \sim (\omega /c)^3$ (in atomic units
in which $m_e=\hbar =e^2=1$ and the speed of light $c\approx 137$).
One suggestion applicable to atoms with positive electron affinities, was
to use a charge-transfer reaction for negative ions,
$A^-+e^+\rightarrow e^+A+e^-$, and measure either its threshold energy or the
electron spectrum \cite{MR99,MI01}. The cross section of this process
should be atomic-sized, but this scheme has not been realized experimentally
yet.

In contrast, much is now known about positron binding to molecules. Binding
energies for over thirty polyatomic species have been determined
\cite{DYS09,DGS10} by measuring positron annihilation with a high-resolution,
tunable, trap-based positron beam \cite{GKG97}. The key idea of this method
is that for molecules that are capable of binding the positron, the dominant
annihilation mechanism is through formation of positron-molecule
vibrational Feshbach resonances \cite{Gr00,Gr01,GYS10}. The majority of
the resonances observed are associated with individual vibrational modes of
the molecule. The binding energy $\eb $ can then be found from the downshift
of the resonance energy
$\eps _\nu =\omega _\nu -\eb $
with respect to the energy $\omega _\nu $ of the vibrational excitation
\cite{GBS02,BGS03}. These experiments proved the link between positron
binding and enhanced annihilation rates \cite{GYS10}.


For atoms existing theoretical predictions of positron binding are limited to
species
with one or two valence $s$ electrons, as these systems are easier to compute.
It is expected that many other atoms with open multielectron valence shells,
can bind the positron \cite{DFG95,MR99}.
Physically, positron binding is facilitated by a sizeable
dipole polarizability $\alpha _d$ and moderate ionization potential $I$.
While there is no rigorous criterion for binding, examination of the atoms
that bind, suggests the following conditions: $\alpha _d\gtrsim 40$~a.u. and
$I<10$~eV.

Large values of $\alpha _d$ ensure that the positron experiences a strong
attractive
polarization potential $-\alpha _d/2r^4$ outside the atom. Small ionization
potentials increase the effect of virtual positronium (Ps) formation:
a process in which an atomic electron temporarily joins the positron. 
It gives a distinct contribution to the positron-atom attraction
akin to covalent bonding \cite{DFG95,ACC76,DFK93}. The energy of the
ground-state Ps is $E_{\rm 1s}=-6.8$~eV, and
this effect is strongest for $I\sim 6.8$~eV. For atoms with $I<|E_{\rm 1s}|$,
positron bound states increasingly have the
character of a ``Ps cluster'' orbiting the positive ion \cite{MBR99}. In this
case the criteria for binding change, atoms with compact cores being favoured 
(e.g., $e^+$Na is bound while $e^+$K is not). Atoms with $I<6.8$~eV also differ
in one other important aspect: the Ps-formation channel
($A+e^+\rightarrow A^++\mbox{Ps}$) is open at all positron energies for them.

\begin{figure}[ht]
\includegraphics*[width=8.5cm]{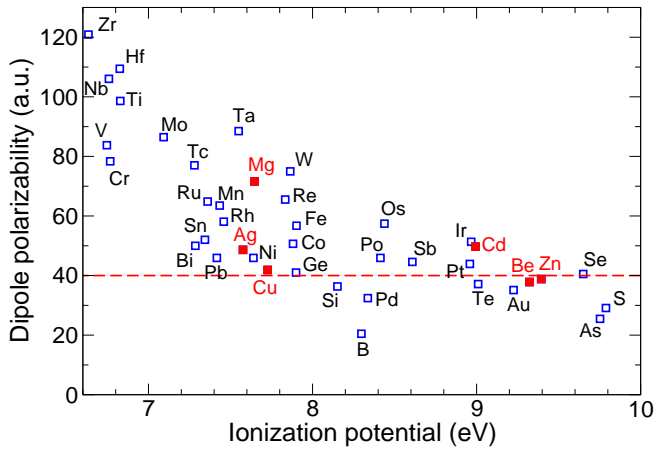}
\caption{Dipole polarizabilities $\alpha _d$ (from Ref. \cite{CRC}, except for
Au \cite{SB94}) vs. ionization potentials for atoms in the range where positron
binding can be expected. Solid squares show atoms for which positron binding
is predicted by high-quality calculations. The horizontal line
$\alpha _d=40$~a.u. is an approximate boundary between binding and non-binding
atoms.}
\label{fig:alphaI}
\end{figure}

Figure \ref{fig:alphaI} shows the polarizabilities vs. ionization potentials
for atoms with $6.6<I<10$~eV. For most of them $\alpha _d>40$~a.u., and
according to the above criterion, they are likely to form bound states with
the positron. Solid symbols identify atoms for which the binding energies have
been calculated: Be, $\eb =87$~meV \cite{BM06}; Zn, 103~meV \cite{MZB08};
Cd, 126~meV \cite{BM10}; Ag, 123~meV \cite{DFH00}; Cu, 170 meV \cite{DFG99};
and Mg, 464~meV \cite{MZB08}. The weakest binding in this group is by Be and
Zn found on the bottom right in Fig. \ref{fig:alphaI}. For a nearby atom of
gold positron binding occurs in the nonrelativistic approximation (which
underestimates the ionization potential and overestimates
$\alpha _d$). However, in a fully relativistic calculation this system is not
bound~\cite{DFH00}.

Figure \ref{fig:alphaI} shows that
most good candidates for positron binding are transition metal atoms
with open $p$ or $d$ subshells. Many of these atoms possess low-lying excited
states with energies $\sim 1$~eV, due to a fine-structure or Coulomb splitting
of the ground-state configuration, or $ns$-$(n-1)d$ transitions. Their
polarizabilities are similar to those of the ground states. Hence, they are
also likely to bind the positron. Depending on the excitation and binding
energies, this will be either a true bound state ($\eb >\omega _\nu $), or a
resonance in the positron continuum ($\eb < \omega _\nu $).



To substantiate the claim that open-shell atoms can bind positrons, we have
estimated their binding energies using many-body theory calculations. In this
approach the positron wavefunction $\psi $ and energy $\eps _0$
are found from the Dyson equation \cite{DFG95},
\begin{equation}\label{eq:Dys}
(H_0+\Sigma )\psi =\eps _0\psi ,
\end{equation}
where $H_0$ is the positron Hamiltonian in the field of the 
ground-state atom in the Hartree-Fock approximation, and $\Sigma $ is 
a nonlocal operator that describes the effect of correlations
(``correlation potential''). In the lowest order of many-body perturbation
theory $\Sigma $ is given by the 2nd-order contribution $\Sigma ^{(2)}$
which accounts for the atomic polarization \cite{GL04}. Higher-order
contributions, such as that
of virtual Ps formation, make the total correlation potential stronger. They
are also more difficult to evaluate. To gauge their importance we solved
Eq. (\ref{eq:Dys}) with $\Sigma =\zeta \Sigma ^{(2)}$ and adjusted
the coefficient $\zeta $ to reproduce the known binding energies
for Be, Mg, Zn, Cd, Cu, and Ag, which yielded $\zeta =2.0$, 2.4, 1.8, 1.7,
2.4, and 1.9, respectively. The advantage of this approximation is that
it allows one to evaluate $\Sigma $ for open-shell atoms and excited states,
by using orbital occupation factors in the calculation of $\Sigma ^{(2)}$.
Estimates of the binding energies for a number of open-shell atoms in the
ground and excited states obtained using $\Sigma =\zeta \Sigma ^{(2)}$ with
$\zeta =2$, are listed in Table \ref{tab:atoms}. They range between 20 meV
and 0.5 eV. We have checked that even for a smaller value,
e.g., $\zeta =1.7$, all the atoms in the table retain binding.

\begingroup
\begin{table*}[ht]
\squeezetable
\caption{Atoms with low-lying excited states in which positron binding and
annihilation resonances are expected.}
\label{tab:atoms}
\begin{ruledtabular}
\begin{tabular}{rllcccllcc}
$Z$ & Atom & Ground & $I$ & $\alpha _d$\footnotemark[1] &
$\eps _b$\footnotemark[2] & Excited & $\eps _b$\footnotemark[3] &
 $\omega _\nu$\footnotemark[4] & Transition \\
 & & state & (eV) & (a.u.) & (eV) & state(s) & (eV) & (eV)
 & type\footnotemark[5] \\
\hline
26 & Fe & $3d^64s^2~^5\mbox{D}_4$ & 7.902 & 56.7 & 0.28 &
$3d^74s~^5\mbox{F}_J$ & 0.09 & 0.859--0.990 & $E2$ \\
27 & Co & $3d^74s^2~^4\mbox{F}_{9/2}$ & 7.881 & 50.7 & 0.26 &
$3d^84s~^4\mbox{F}_J$ & 0.08 & 0.432--0.582 & $E2$ \\
28 & Ni & $3d^84s^2~^3\mbox{F}_4$ & 7.640 & 45.9 & 0.24 &
$3d^84s^2~^3\mbox{F}_2$ & 0.24 & 0.275 & $E2$ \\
& & & & & &
$3d^94s~^3\mbox{D}_1$ & 0.07 & 0.212 & $E2$ \\
& & & & & &
$3d^94s~^1\mbox{D}_2$ & 0.07 & 0.423 & $E2$ \\
43 & Tc & $4d^55s^2~^6\mbox{S}_{5/2}$ & 7.280 & 77.0 & 0.46 &
$4d^65s~^6\mbox{D}_J$  & 0.23 & 0.319--0.518 & $E2$ \\
44 & Ru & $4d^75s~^5\mbox{F}_5$ & 7.361 & 64.9 & 0.21 &
$4d^75s~^5\mbox{F}_J$  & 0.21 & 0.259--0.385 & $E2,M3,E4$ \\
45 & Rh & $4d^85s~^4\mbox{F}_{9/2}$ & 7.459 & 58.1 & 0.20 &
$4d^85s~^4\mbox{F}_J$ & 0.20 & 0.322,~0.431 & $E2,M3$ \\
& & & & & &
$4d^9~^2\mbox{D}_J$ & 0.10 & 0.410,~0.701 & $E2,M3$ \\
& & & & & &
$4d^85s~^2\mbox{F}_{7/2}$ & 0.20 & 0.706 & $E2$ \\
50 & Sn & $5s^25p^2~^3\mbox{P}_0$ & 7.344 & 52.0 & 0.02 &
$5s^25p^2~^3\mbox{P}_J$  & 0.02 & 0.210,~0.425 & $M1,E2$ \\
51 & Sb & $5s^25p^3~^4\mbox{S}_{3/2}$ & 8.608 & 44.6 & 0.05 &
$5s^25p^3~^2\mbox{D}_J$  & 0.05 & 1.055,~1.222 & $E2$ \\
73 & Ta & $5d^36s^2~^4\mbox{F}_{3/2}$ & 7.550 & 88.5 & 0.45 &
$5d^36s^2~^4\mbox{F}_J$  & 0.45 & 0.249,~0.491 & $E2$ \\
74 & W  & $5d^46s^2~^5\mbox{D}_0$ & 7.864 & 75.0 & 0.46 &
$5d^46s^2~^5\mbox{D}_J$ & 0.46 & 0.209--0.771 & $M1,E2,M3,E4$ \\
& & & & & &
$5d^56s~^7\mbox{S}_3$ & 0.30 & 0.366 & $M3$ \\
& & & & & &
$5d^46s^2~^2\mbox{P}_0$ & 0.46 & 1.181 & $E0$ \\
76 & Os & $5d^66s^2~^5\mbox{D}_4$ & 8.438 & 57.4 & 0.47 &
$5d^66s^2~^5\mbox{D}_J$ & 0.47 & 0.340--0.755 & $E2,M3,E4$ \\
& & & & & &
$5d^76s~^5\mbox{F}_J$ & 0.29 & 0.638--1.614 & $E2,M3$ \\
& & & & & &
$5d^66s^2~^3\mbox{P}_2$ & 0.47 & 1.260 & $E2$ \\
 & & & & & &
$5d^76s~^3\mbox{F}_J$ & 0.29 & 1.368,~1.747 & $E2$ \\ 
77 & Ir & $5d^76s^2~^4\mbox{F}_{9/2}$ & 8.967 & 51.3 & 0.46 &
$5d^76s^2~^{2S+1}L_J$ & 0.46 & 0.506--2.204
& $E2,M3,E4$ \\
& & & & & &
$5d^86s~^{2S+1}L_J$ & 0.28 & 0.351--2.068
& $E2,M3$ \\
78 & Pt & $5d^96s~^3\mbox{D}_3$ & 8.960 & 43.9 & 0.27 &
$5d^96s~^3\mbox{D}_J$ & 0.27 & 0.814,~1.256 & $E2$  \\
& & & & & &
$5d^86s^2~^{2S+1}L_J$ & 0.46 & 1.254--2.106 & $E2,M3$ \\
& & & & & &
$5d^{10}~^1\mbox{S}_0$ & 0.23 & 0.761 & $M3$ \\
\end{tabular}
\footnotetext[1]{Dipole polarizabilities from Ref. \cite{CRC}.}
\footnotetext[2]{Binding energies $\eb =|\eps _0|$ for atoms in ground-state
configurations obtained using $\Sigma =\zeta \Sigma ^{(2)}$ with
$\zeta =2$ (see text).}
\footnotetext[3]{Binding energies for atoms in excited-state configurations
obtained with $\Sigma =\zeta \Sigma ^{(2)}$, $\zeta =2$.}
\footnotetext[4]{Energies of excited states from Ref. \cite{NIST}, such that
$0.2~\mbox{eV} < \omega _\nu<E_{\rm Ps}+0.15$~eV.}
\footnotetext[5]{When several transitions are allowed, the most probable
is indicated.}
\end{ruledtabular}
\end{table*}
\endgroup

Table \ref{tab:atoms} also lists the energies of low-lying atomic excited
states. The state of a positron bound to an {\em excited} atom, which lies
above the atomic ground state, is a Feshbach resonance. Its total width
$\Gamma $ is determined by the annihilation width $\Gamma ^a$ and elastic
width $\Gamma ^e$. The latter gives the decay rate of this quasibound state
into the $A+e^+$ continuum, and also characterizes the probability of its
formation in positron-atom collisions. Strong annihilation
resonances require $\Gamma ^e\gg \Gamma ^a$ (see below).

An additional consideration used in selecting the atoms in
Table \ref{tab:atoms}, was the energy of their Ps-formation threshold
$E_{\rm Ps}=I-|E_{\rm 1s}|$. If a positron resonance lies above
$E_{\rm Ps}$, it can also decay into the $A^++\mbox{Ps}$ channel.
This can significantly increase the total width, reducing the size of the
annihilation resonance. More importantly, for incident positron energies
$\eps >E_{\rm Ps}$, Ps formation becomes the dominant annihilation channel,
and the resonant annihilation signal is ``drowned''. Hence, we focus on
resonances that lie {\em below} the Ps-formation threshold.

The last column in Table~\ref{tab:atoms} indicates the type of electromagnetic
transitions between the ground and excited states allowed by selection
rules. All of the excited states have the same parity as the ground
state. The majority of them have nonzero total angular
momenta $J$. As a result, the most common allowed transition between the
levels is $E2$. Of course, the excitation of an atom by the Coulomb field of
the positron in the process of capture, is different from that by a photon.
However, for the transitions of electric type, a simple estimate of the
elastic width in terms of the atomic transition amplitude can be derived,
which shows that $\Gamma ^e\gg \Gamma ^a$ (see below).

The resonant contribution to the positron-atom annihilation cross section
is written using the Breit-Wigner formalism \cite{Landau} as
\begin{equation}\label{eq:BW}
\sigma _a=\frac{\pi }{k^2}\sum _\nu \frac{2J_\nu +1}{2J+1}\,
\frac{\Gamma ^a_\nu \Gamma ^e_\nu }{(\eps -\eps _\nu)^2+\Gamma _\nu ^2/4},
\end{equation}
where $k=\sqrt{2 \eps }$ is the positron momentum, $J$ is the total angular
momentum of the target ground state, and $J_\nu $ is that of the resonance
$\nu $. To estimate the observable effect, we average the normalized
dimensionless annihilation rate $\Z =\sigma _ak/(\pi r_0^2c)$ ($r_0$ being the
classical electron radius) over the energy distribution in the positron beam,
and obtain
\begin{equation}\label{eq:Zeff}
\Z (\epsilon ) =\frac{2\pi ^2\rho _{ep}}{2J+1}\sum _\nu
\frac{(2J_\nu +1)\Gamma ^e_\nu }{k_\nu\Gamma _\nu }\Delta (\epsilon -\eps_\nu ),
\end{equation}
where $\epsilon $ is the mean longitudinal energy of the beam,
$k_\nu =\sqrt{2\eps _\nu}$, and $\rho _{ep}$ is the electron-positron contact
density in the positron bound state, which determines its annihilation width
$\Gamma ^a_\nu =\pi r_0^2c\rho _{ep}$. The function
$\Delta $ describes the positron energy distribution around the mean energy
$\epsilon $, $\int \Delta (E)dE=1$ \cite{GL06,YGL08}.

The contact density can be estimated from
$\rho _{ep}\approx (F/2\pi )\sqrt{2\eb }$, where $F=0.66$ \cite{Gr01,GYS10}.
To evaluate the elastic width, we use a multipole
expansion of the positron Coulomb interaction with the atom.
Using the fact that for a low-energy positron, large positron-atom
separations dominate, one obtains (cf. Ref. \cite{GL06}),
\begin{equation}\label{eq:Game}
\Gamma ^e_\nu =\frac{8 k _\nu^{2\lambda } 
|\langle \nu J_\nu \|Q_\lambda \|0J\rangle |^2}
{(2\lambda +1)(2J_\nu+1)[(2\lambda +1)!!]^2}\,f_\lambda (\eps _\nu/\eb ),
\end{equation}
where $\lambda $ is the order of the multipole (e.g., $\lambda =2$
for a quadrupole excitation), $\langle \nu J_\nu \|Q_\lambda \|0J\rangle $
is the reduced matrix element of the $2^\lambda $-pole moment
between the ground (0) and excited ($\nu $) atomic states, and
$f_\lambda (x)=\sqrt{x/(1-x)}
\left[ _2F_1(\frac{1}{2},1;\lambda +\frac 32;-x) \right]^2$
is a dimensionless function, such that $f_\lambda (x)\sim 1$ for $x\sim 1$.

Estimating $\Gamma ^e _\nu $ from Eq.~(\ref{eq:Game}) for a Feshbach resonance
at $\eps _\nu \sim 1$~eV, populated through a quadrupole transition, and
assuming that the quadrupole amplitude is atomic-sized, one obtains
$\Gamma ^e _\nu \sim 1$--10~meV. Hence, this resonances are sufficiently narrow
to produce observable sharp features in the energy dependence of $\Z$.
Estimating the annihilation width for $\eb = 150$~meV, we obtain
$\Gamma ^a_\nu = 4\times 10^{-7}$~eV, hence, $\Gamma ^e _\nu \gg \Gamma ^a_\nu$.
In this case, $\Gamma ^e_\nu /\Gamma _\nu \approx 1$, and the contribution
of such resonance to $\Z$, Eq.~(\ref{eq:Zeff}), is close to maximum possible.
For a positron beam with energy spread $\delta \eps \sim 25$~meV,
using $\Delta _{\rm max}\sim 1/\delta \eps $, the peak resonant value of
the annihilation rate from Eq.~(\ref{eq:Zeff}) is given by
$\Z \sim \pi F\sqrt{\eb /\eps _\nu }/\delta \eps \sim 10^3$.
This estimate remains valid even if the elastic width is suppressed by up
to three orders of magnitude, e.g., for a higher-multipole transition, or a
transition mediated by the relativistic (spin-orbit) interaction.


The above analysis indicates that positron-atom resonances could be observed
with a trap-based beam used for studying resonances in positron-molecule
annihilation \cite{BGS03}. Such a measurement requires vapour pressure
of $\sim 0.01$~mtorr \cite{GBS02}. For the atoms in Table \ref{tab:atoms}
this can be achieved be heating the samples to temperatures ranging from
650\,$^\circ$C for Sb and 1100\,$^\circ$C for Sn, to $1500$\,$^\circ$C for
Fe, Co, and Ni, and over 2000\,$^\circ$C for other species \cite{vapor}.
Detection of the resonances can thus provide the first experimental evidence
of positron binding to neutral atoms, and first estimates of the binding
energies. While a Feshbach resonance only signifies binding to an excited
state, the binding energy in the ground state is expected to be similar
if it has the same electronic configuration.

Resonant enhancement can also be observed with thermalized
positrons. Depending on the exact position of the resonances, it can lead to
a nontrivial dependence of the annihilation rate on the positron temperature,
with greater rates measured at higher temperatures. Such behaviour would be
in sharp contrast with that observed in nonresonant systems, such as the noble
gases \cite{KGS96}.


One should mention that earlier experimental searches for positron resonances
in the vicinity of electronic excitation thresholds (for H$_2$, N$_2$, CO and
Ar) yielded negative results \cite{SGB01}. However, these systems are quite
different from the open-shell metal atoms considered here. None of them is
expected to bind the positron in the ground state, and the electronic
excitations lie above the Ps formation threshold. In addition, the relative
role of resonances in the annihilation is much more prominent than in the
elastic or total scattering measured in Ref.~\cite{SGB01}.

The case of transition-metal atoms is also markedly different from that
of Be, in which positron binding to the excited $2s2p~^3\mbox{P}$
state was predicted in configuration-interaction calculations \cite{BM07}.
This excited state lies above the Ps formation threshold, but a large
positron binding energy of $250$~meV ensures that the bound state
is $40$~meV below the $\mbox{Be}^++\mbox{Ps}$ threshold. Such strong
binding by the excited state is promoted by its large dipole polarizability.
For comparison, the positron binding energy by the the ground-state Be atom
is 87~meV \cite{BM06}.

Besides the Feshbach resonances, positron annihilation can be increased by
shape resonances. These resonances are supported by the strong polarization
attraction and the centrifugal barrier. Thus, calculations predict a sharp
$p$-wave
resonance in positron scattering from Mg at 95~ meV, with $\Z=1300$ at the
peak, and similar but broader resonances at $\sim 0.45$--0.65~eV in Cu, Zn,
and Cd, with $\Z \sim 100$ \cite{MZB08,BM10}. Compared with the Feshbach
resonances, the shape resonances do not indicate positron binding. They
also have much larger widths, e.g., $\sim 0.1$~eV in Mg.



\begin{acknowledgments}
The authors acknowledge useful discussions with C.~M.~Surko. This work was
supported by the Australian Research Council and Gordon Godfrey fund.
\end{acknowledgments}

\end{document}